\documentclass[12pt,preprint]{aastex}
\slugcomment{submitted to Ap.J.}
\shortauthors{Dhawan et al.}
\shorttitle{Kinematics of Black Hole in GRS~1915+105}

\begin{document}

\title{Kinematics of Black Hole X-ray Binary GRS~1915+105}

\author{V. Dhawan}
\affil{National Radio Astronomy Observatory, Socorro, NM 87801 USA.}
\email{vdhawan@nrao.edu}

\author{I.~F. Mirabel \altaffilmark{1} } 
\affil{European Southern Observatory, Casilla 19001, Santiago 19,
       Chile}
\email{fmirabel@eso.org}
\altaffiltext{1}{\footnotesize On leave from Service d'Astrophysique,
CEA-Saclay, 91191 Gif-sur-Yvette, France.}

\author{M. Rib{\'o}}
\affil{Departament d'Astronomia i Meteorologia, Universitat de 
       Barcelona, \\ Marti i Franques 1, 08028 Barcelona, Spain }
\email{mribo@am.ub.es}

\author{I. Rodrigues}
\affil{Universidade do Vale do Paraiba, Sao Jose dos Campos, Brazil}
\email{irapuan@if.ufrgs.br}

\begin{abstract}

The space velocity of a stellar black hole encodes the history of its
formation and evolution.  Here we measure the 3-dimensional motion of
the microquasar GRS~1915+105, using a decade of astrometry with the
NRAO Very Long Baseline Array, together with the published radial
velocity. The velocity in the Galactic Plane deviates from circular
rotation by 53--80$\pm$8~km~s$^{-1}$, where the range covers any
specific distance from 6--12~kpc. Perpendicular to the plane, the
velocity is only 10$\pm$4~km~s$^{-1}$. The peculiar velocity is
minimized at a distance 9--10~kpc, and is then nearly in the radial
direction towards the Galactic Center.  We discuss mechanisms for the
origin of the peculiar velocity, and conclude that it is most likely a
consequence of Galactic velocity diffusion on this old binary, rather
than the result of a supernova kick during the formation of the
14~M$_{\odot}$ black hole. Finally, a brief comparison is made with 4
other BH binaries whose kinematics are well determined.

\end{abstract}


\section{Introduction}

   GRS~1915+105 is a famous microquasar with superluminal ejecta
\citep{MR94}, and spectacular X-ray variability from an unstable
accretion disk \citep{GMR96, Bel00}.  For nearly a decade, its nature was
highly obscured at a distance $D$ of $\sim$10~kpc.  Finally, IR
spectroscopy with the VLT \citep{GCM01} was used to measure the
heliocentric radial velocity of $-$3$\pm$10 km s$^{-1}$ for the center
of mass of the binary, a mass of 14$\pm$4~M$_{\odot}$ for the compact
object, with a donor mass of $\sim$1.2$\pm$0.2~M$_{\odot}$ in an orbit
of 33.5~days period. VLBA imaging \citep{Dha00a}, in addition to the
ejecta during outbursts, revealed an AU-sized radio jet closely
coupled to the accretion disk activity. Under the conditions imaged
here, the so-called low-hard state, the flux of the AU-size jet
follows the variation in X-rays with a time-lag of a few minutes,
\citep{Metal98, Dha00a}.  The jet is hence located within 1~AU
(0.1~mas) of the black hole.

\section{Observations and Results}

\subsection{Observations}

  Our observations spanned $\sim$10 years, sampled at random times
dictated by the flare activity of the source. Usually the goal was to
image the AU-size jet and larger scale superluminal ejecta in
coordination with X-ray instruments. The moving ejecta can confuse the
position determination. Indeed there were some occasions just after a
major flare when the central nucleus was very weak or absent, and the
moving ejecta dominated the flux density of the source.  We have hence
selected for this astrometric study only those images which were
dominated by the central source of flux density 10 to 50~mJy,
corresponding to the AU-size flat-spectrum nuclear jet. Examples of
images are published in \citet{Dha00a, Fuc03, RDM04, MJ07}.

Standard VLBA phase-referencing was used, \citep{BC95} with a total
 cycle time of 3~minutes, of which 1~minute was on the primary
 calibrator J1924+1540, at a separation of 5.3$^{\circ}$ from the
 target.  This technique provides the position of the black hole
 relative to the reference source, whose position has been
 pre-determined in the ICRF frame.  Most of the observations were at a
 frequency of 8.4 GHz (3.6cm wavelength). During some observations,
 depending on the radio spectra, (if available from monitoring with
 other instruments) we observed at 3.6 and 2~cm. These were alternated
 every 20min, at 256Mbit~s$^{-1}$ recording.  Standard post-processing
 corrections were applied for ionospheric refraction using a GPS-based
 global model. A-priori amplitude calibration was done from measured
 system temperatures and antenna gains, followed by self-calibration
 on the reference source, application of the refined calibration to
 all sources including the target, and imaging.

We fitted a single elliptical Gaussian to each image to extract the
 position. The formal errors of the fit are far exceeded by
 tropospheric variations, resulting in position errors $\sim$1~mas
 {\it rms} at each epoch. The earliest observations have larger error
 due to the poor choice of reference sources, J1922+1530 (4.8
 $^{\circ}$ away, but quite resolved) and J1925+2106, 10.5$^{\circ}$
 away from the target.  We estimated the errors, including systematic
 errors peculiar to the weather on that day, from the scatter in image
 position over intervals of $\sim$20~min. In later observations, a
 second calibrator, J1922+0840 (2.9 $^{\circ}$ away), was observed as
 a quality check and processed in identical fashion to the target.
 Positions at the two wavelengths on a given observation date agreed,
 and were averaged together for that date. On one occasion
 observations were possible on successive days, positions from which
 were averaged together.

\subsection{Results}

The measurements \footnotemark{} 
\footnotetext{\footnotesize Strictly, the measured VLBA positions are
Geocentric. In the table, plot \& fit, we report measured quantities,
rather than correcting them for parallax with an uncertain distance.
In the fit for proper motion, the error is reduced by the random
observation date, and the long timespan.  The parallax is 0.1~mas at
$D$=10~kpc. The changes in proper motion if we correct for parallax are
$\sim$0.01~mas~yr$^{-1}$, far below our measurement errors. Thus
Geocentric positions can safely be taken as Heliocentric.}
are summarized in Table-{\ref{tab1} and Figure-{\ref{fig1}, showing
the coordinates of the black hole versus time. We fit the following
position and proper motion at MJD 51545 (01 January 2000):

RA~(J2000) 19$^{\rm h}$15$^{\rm m}$11$^{\rm s}$.54902 $\pm$0.7~mas
~~~~Dec(J2000) +10$^{\circ}$56'44''.7478 ~$\pm$0.9~mas

$\mu_{\rm RA}$ Cos(Dec) = $-$2.86 $\pm$0.07~mas~yr$^{-1}$
~~~~$\mu_{\rm Dec}$     = $-$6.20 $\pm$0.09~mas~yr$^{-1}$

The corresponding Heliocentric motions in Galactic coordinates are:

$\mu_{\it l}$ = $-$6.82 $\pm$0.10~mas~yr$^{-1}$
~~~~$\mu_{\it b}$ = $-$0.35 $\pm$0.07~mas~yr$^{-1}$

Using our measured proper motion and the known V$_{\rm radial}$ of
$-$3$\pm$10~km~s$^{-1}$, we transform these measured Heliocentric
quantities with the standard method of \citet{JS87} updated for J2000,
and the Solar peculiar motion measured by Hipparcos, [U$_{\odot}$,
V$_{\odot}$, V$_{\odot}$ = 10.0, 5.25, 7.17 km~s$^{-1}$] from
\citet{DB98}. We use the convention U + toward {\it l} = 0$^{\circ}$;
V + toward {\it l} = 90$^{\circ}$; W + toward the North Galactic Pole.

We compare these measured velocities with those expected from circular
rotation at 220~km~s$^{-1}$ at the location of the black hole.  For
the distance, 11 to 12~kpc have been proposed, \citep{Fen03, Dha00b},
but since considerably smaller values are suggested by others, e.g.
\citet{Kai04}, we let the distance be a free parameter in
Figure~\ref{fig2} and Table~\ref{tab2}.  We then evaluate the velocity
discrepancy in local and Galactocentric frames, and note that:

\begin{itemize}

\item The measured W component is small, so the black hole velocity
lies primarily in the Galactic Plane.

\item At any distance there is a significant component of
measured velocity directed inward to the Galactic Center, whereas none
is expected for purely circular motion.

\item The extremum (negative) proper motion from Galactic rotation
  along this line of sight occurs at $D$=~9~kpc, and is
  $-$5.85~mas~yr$^{-1}$, while $-$6.82~mas~yr$^{-1}$ is measured.

\item Figure \ref{fig3} shows that the total peculiar velocity of the
black hole (in either LSR or GC frame) is minimized at $D$=~9--10~kpc,
and is then 53~km~s$^{-1}$ in the radial direction towards the
Galactic Center.

\item The measured heliocentric velocity matches that expected from
circular rotation at $\sim$12~kpc, as already noted by \citet{GCM01}.
However, at this distance, the total peculiar velocity is
83~km~s$^{-1}$.

\item At $D$=~6.5~kpc, both the Galactocentric components mismatch
circular rotation by over 45~km~s$^{-1}$ each.

\end{itemize}

Hence, the distance favoured by velocity matching is 9--10~kpc, but
this is not a very strong argument. A distance determined by
trigonometric parallax, while possible with the VLBA, remains very
challenging for this object.

\subsection{The Galactocentric Orbit}

We next compute the Galactocentric orbit of the black hole assuming a
 smooth Galactic potential composed of disk, spheroid and halo,
 following \citet{JSH95}. Figure~\ref{fig4} shows the orbit assuming a
 distance of 12~kpc. The path for the past 1~Gyr is the blue (dashed)
 line, with the last 250~Myr in red (solid line).  The orbit deviates
 from circularity with an eccentricity of 0.285, and reaches a maximum
 height above the Galactic Plane of only 230~pc, typical of the thin
 disk population. However, GRS~1915+105 does not match the velocity of
 thin disk objects at any distance.  Detailed computation of the orbit
 including the effects of spiral arms is beyond the scope of this
 paper.

\section{Discussion}

We discuss mechanisms by which the black hole could have acquired its
anomalous velocity. To summarize, though the velocity is significant,
its component out of the Galactic plane is low, so an impulsive origin
cannot be proved. The age and uncertain distance of the system
disallow a strong conclusion about the origin of the velocity.
However, our data helps to rank the probability of three mechanisms.

\subsection{Stellar Diffusion}

An individual star in the disk can acquire significant changes in
angular momentum from non-axisymmetric forces such as spiral arms,
leading to an increase of its non-circular motion with age
\citep{Wei77, SB02}.  If we assume the binary is as old as the donor,
a red giant of 1.2~M$_{\odot}$ aged $\sim$4~Gyr, then it has endured
about 16 Galactic orbits. Velocity imparted by spiral arm passages
conforms naturally to the measured low velocity perpendicular to the
Galactic plane, and has had sufficient time to operate, given the age
of the system. In comparison, the non-circular velocity of
progressively older disk stars in the solar neighborhood is
30--50km~s$^{-1}$, as discussed for example in \citet[][Table
6]{Mig00}, and \citet[][population components S$_{3}$ and
A$_{3}$]{Alc05}.  At a distance $D\sim$9--10~kpc, the above mechanism
suffices to account for the peculiar velocity, and must have operated
to some extent.  It may have difficulty in producing a velocity
perturbation of 70--80~km~s$^{-1}$, as would be the case if the source
were at 6 or at 12~kpc.

\subsection{Natal Kicks}

We note that even at 9~kpc, the black hole possesses the momentum of a
pulsar of 1.4~M$_{\odot}$ with velocity of 530~km~s$^{-1}$.  Hence in
the following we discuss the possible origin of the peculiar velocity
in a natal supernova event, assuming one occurred. We examine this
assumption in a later section.

\subsubsection{Blaauw Kick} 

Mass unbound suddenly and symmetrically by the supernova explosion
forces the binary to recoil with a momentum opposite to that of the
unbound mass at the time of ejection. This spherical mass loss results
in a kick that lies in the plane of the binary orbit.  We estimate,
using equation~(1) of \citet{Nel99}, that mass unbound from the binary
in a SN explosion imparted at most 23--44~km~s$^{-1}$, for allowable
mass ratios as measured by \citet{GCM01}.  Assuming the binary plane
was randomly oriented, the kick has $\sim$25\% probability of lying in
the galactic plane within $\pm$15$^{\circ}$, which is the ratio of
measured peculiar velocities perpendicular and parallel to the plane.

\subsubsection{Asymmetric Kick:} 

 If the binary lies at $D$=~12~kpc or 6~kpc, an {\em additional} kick
of $\sim$30~km~s$^{-1}$ is required. It is generally accepted that
neutron stars form in asymmetric SN, and receive substantial birth
kicks, which are seen in the velocities of young single pulsars,
$\sim$400km~s$^{-1}$ \citep{Hob05}. Neutron stars in long period
binaries may receive smaller kicks of $\sim$50~km~s$^{-1}$
\citep{Pfa02}, which is a case more closely related to the present
binary with orbital period 33~days. This kick due to any asymmetries
in the explosion has no preferred direction relative to the binary
orbit plane, unless it is assumed that the progenitor's spin is
parallel to the orbital axis.  Assuming random orientation, an
asymmetric supernova has roughly the same 25\% chance (as above), of
producing a kick that lies in the Galactic plane.

\subsection{Comparison with the kinematics of other black hole binaries}

The question of whether or not black holes receive a natal kick
similar to neutron stars is an open one.  On the theoretical side, the
formation of black holes of over 10~M$_{\odot}$ challenges stellar
evolution models. \citet{Fry01} proposed that black holes of that mass
should form by direct collapse without SN explosion, or after a SN
with energy insufficient to unbind the stellar envelope.  \citet{BB02}
also conclude that standard models would work only if wind mass loss
for very massive stars was half as much as commonly assumed, and black
holes of $\ge$10~M$_{\odot}$ would form by direct collapse.  Hence no
kick velocity is expected in the formation of the most massive BH.
There is some observational evidence that energetic SN may not occur
in the formation of stellar black holes of relatively high mass.
Cyg~X-1 is a 10~M$_{\odot}$ BH with a peculiar velocity of
$<$10~km~s$^{-1}$, and may be an example of such a `dark birth'
\citep{MR03a}.

Less massive black holes can form through partial fallback onto the
collapsing object, after a SN explosion. We briefly evaluate the
evidence for runaway black holes which have well measured,
3-dimensional velocities.  The most likely mechanism for their high
velocity is an asymmetric kick from a SN. Independent confirmation
comes from rather difficult measurement of the supernova's
nucleosynthesis products in the companion star. Such evidence exists
for neutron stars as well, e.g., \citet{GH05}.

The runaway binary ~GRO~1655$-$40 contains a dynamically confirmed BH
of 6.3~M$_{\odot}$ \citep{Gre01}, and has an anomalous velocity
v=~120~km~s$^{-1}$ \citep{Metal02}. \citet{Isr99} found super-solar
abundances of several $\alpha$-elements in the atmosphere of the
companion star, and concluded that SN ejecta had enriched the
companion.  \citet{Foe07}, using different techniques, found little
evidence for enrichment except in O.  More recently, \citet{GH07},
using VLT/UVES spectra, support the abundances derived by
\citet{Isr99}.

~XTE~1118+48 is another confirmed 7--8~M$_{\odot}$ BH binary
\citep{Gel06}, orbiting in the Galactic halo with
v=~145~km~s$^{-1}$. \citet{Metal01} proposed that it was born in a
globular cluster $\sim$7~Gyr ago. However, more recently \citet{Gon06}
claim to have found fossil evidence of a supernova in the metal
enrichment of the small mass donor, and propose instead that the
progenitor was born in the Galactic disk and the binary shot out to
the halo by the SN explosion.

LS~5039 has a runaway velocity of 126~km~s$^{-1}$, \citep{Rib02}, and
has clearly been kicked, but its mass estimate is closer to the
neutron star range than the objects discussed above. It may contain a
black hole of 3.7M$_{\odot}$ \citep{Cas05}, if the binary is
synchronized at periastron. Without the assumption of
pseudo-synchronization, the mass might be as low as 1.4M$_{\odot}$.
Regarding the direct detection of SN byproducts, \citet{McS04} find N
enrichment, but they caution that it could result from internal mixing
due to rapid rotation, a scenario also favoured by \citet{Cas05}.

Finally, GRS~1915+105 does not fit easily in either the high or low
speed category.  The velocity is significantly higher than
Cyg~X-1. However, we cannot conclude that GRS~1915+105 is a runaway,
because its low velocity out of the plane, and the operation of
velocity diffusion in the plane, imply that any natal kick (symmetric
and/or asymmetric) was probably smaller than $\sim$30~km$^{-1}$.  We
hope our data will permit an analysis such as done by \citet{Wil05}
for GRO~J1655$-$40, who used all available information to reconstruct
the evolution back to the time of core collapse and compact object
formation. The observations, even for the small sample so far, imply a
variety of formation scenarios for black holes.

\section{Conclusions}

\begin{enumerate}

\item The most massive known stellar black hole moves at
(53--80)$\pm$8~km~s$^{-1}$ with respect to its regional standard of
rest, for any specific distance in the range 6--12~kpc.

\item At the distance of 9--10~kpc, the peculiar velocity is a
minimum. If GRS~1915 is 3--5~Gyr old, it could have acquired this
velocity by dynamic diffusion, i.e., non-circular motion imparted by
perturbations such as spiral arms over time.  In this case the black
hole in GRS~1915+105 would have formed without a natal kick, like the
10~M$_{\odot}$ black hole in Cyg~X-1.

\item If the distance is smaller (6~kpc) or larger (12~kpc), then some
additional $\sim$30~km~s$^{-1}$ is required from another mechanism
such as a natal supernova explosion, and the kick is constrained to
lie within $\pm$15$^{\circ}$ of the Galactic plane.

\end{enumerate}

\acknowledgements

 The National Radio Astronomy Observatory is a facility of the
National Science Foundation operated under cooperative agreement by
Associated Universities, Inc. \\
V.D. acknowledges support from the Science Visitor's Programme of the
European Southern Observatory, Chile.  ~~M.R. ackowledges financial
support from the Spanish Ministerio de Educaci\'on y Ciencia through a
\emph {Juan de la Cierva} fellowship linked to the project
AYA2004-07171-C02-01, partially supported by FEDER funds. ~~We thank
Luis F. Rodriguez for helpful comments.  ~~This research has made use
of NASA's Astrophysics Data System and the {\bf astro-ph} preprint
service.

\clearpage

\begin{deluxetable}{llllll}

\tablewidth{0pt}
\tablenum{1}
\tablecolumns{6}
\tabletypesize{\small}
\tablecaption{ICRF Position of  Nucleus of GRS~1915+105.\label{tab1}}
\tablehead{
  \colhead{MJD              } &
  \colhead{RA               } &
  \colhead{$\sigma _{\rm RA}$   } &
  \colhead{Dec              } &
  \colhead{$\sigma _{\rm Dec}$  } &
  \colhead{Comment          }      } 
\startdata
50227  & 0.54975 &  1.8  & 0.7685  & 1.8 &  1996 May 24 /  8.4 GHz \\
50229  & 0.54970 &  1.8  & 0.7706  & 1.8 &  1996 May 26 /  8.4 GHz \\
50299  & 0.54965 &  1.5  & 0.7670  & 1.5 &  1996 Aug 03 /  8.4 GHz \\
50583  & 0.54955 &  1.5  & 0.7634  & 1.5 &  1997 May 15 /  8.4, 15 GHz \\
50744  & 0.54943 &  1.5  & 0.7621  & 1.5 &  1997 Oct 23 /  15 GHz \\
50752  & 0.54945 &  1.5  & 0.7629  & 1.5 &  1997 Oct 31 /  8.4 GHz \\
50913  & 0.54939 &  1.0  & 0.7589  & 1.0 &  1998 Apr 11 /  15 GHz \\
50935  & 0.54937 &  1.0  & 0.7582  & 1.0 &  1998 May 02 /  15 GHz \\
51331  & 0.54906 &  1.0  & 0.7522  & 1.0 &  1999 Jun 02 /  8.4, 15 GHz \\
51338  & 0.54905 &  1.0  & 0.7523  & 1.0 &  1999 Jun 09 /  8.4 GHz \\
52107  & 0.54870 &  0.8  & 0.7384  & 0.8 &  2001 Jul 17,18 / 8.4,15 GHz \\
52722  & 0.54846 &  0.5  & 0.7269  & 0.7 &  2003 Mar 24 /  8.4, 15 GHz \\
52731  & 0.54843 &  0.7  & 0.7265  & 0.9 &  2003 Apr 02 /  8.4, 15 GHz\\
52748  & 0.54840 &  0.5  & 0.7264  & 0.7 &  2003 Apr 19 /  8.4, 15 GHz \\
53794  & 0.54778 &  0.7  & 0.7093  & 0.8 &  2006 Feb 28 /  8.4 GHz \\
53800  & 0.54777 &  0.7  & 0.7095  & 0.8 &  2006 Mar 06 /  8.4 GHz \\
\enddata
\tablecomments{- \\
RA(J2000) offset in seconds from 19$^{\rm h}$15$^{\rm m}$11$^{\rm s}$.\\
$\sigma _{\rm RA}$ in milliarcseconds, i.e. multiplied by Cos(Dec) \\
Dec(J2000) offset from 10$^{\circ}$56'44'' \\
$\sigma _{\rm Dec}$ in milliarcseconds. \\
}
 
\end{deluxetable}

\clearpage

\begin{deluxetable}{rrrrrrrrrrrrrr}

\tablewidth{0pt}
\tablenum{2}
\tablecolumns{14}
\tabletypesize{\footnotesize}
\tablecaption{Velocity Components of GRS~1915+105 vs. Distance.\label{tab2}}
\tablehead{
  \colhead{D            } &
  \colhead{U            } &
  \colhead{$\sigma_{U}$ } &
  \colhead{V            } &
  \colhead{$\sigma_{V}$ } &
  \colhead{W            } &
  \colhead{$\sigma_{W}$ } &
  \colhead{UC           } &
  \colhead{VC           } &
  \colhead{VR$_{H}$     } &
  \colhead{VT$_{H}$     } &
  \colhead{VR$_{G}$     } &
  \colhead{VC$_{G}$     } &
  \colhead{V$_{P} $     }      }
\startdata
5  & 122.9 & 7.5 & $-$110.5 & 7.6 & $-$1.1  & 1.8 & 127.8 & $-$40.9  & 49.9  & $-$116.2 & $-$36.4 & 160.5 & 69.8 \\
6  & 145.9 & 7.6 & $-$133.2 & 7.7 & $-$2.8  & 2.1 & 155.3 & $-$64.2  & 52.7  & $-$152.2 & $-$42.1 & 164.5 & 69.7 \\
7  & 168.9 & 7.7 & $-$155.9 & 7.7 & $-$4.4  & 2.5 & 178.6 & $-$91.6  & 49.6  & $-$188.0 & $-$46.6 & 174.5 & 65.2 \\
8  & 191.9 & 7.7 & $-$178.6 & 7.8 & $-$6.1  & 2.8 & 196.3 & $-$120.7 & 41.3  & $-$221.1 & $-$49.7 & 189.9 & 58.4 \\
9  & 214.9 & 7.8 & $-$201.4 & 7.9 & $-$7.7  & 3.2 & 208.3 & $-$149.2 & 29.4  & $-$249.6 & $-$51.5 & 209.5 & 53.2 \\
10 & 237.9 & 7.9 & $-$224.1 & 8.0 & $-$9.4  & 3.5 & 215.4 & $-$175.4 & 15.8  & $-$273.1 & $-$52.3 & 232.1 & 54.5 \\
11 & 260.9 & 8.1 & $-$246.8 & 8.1 & $-$11.1 & 3.9 & 218.9 & $-$198.4 &  1.9  & $-$291.8 & $-$52.3 & 257.0 & 65.0 \\
12 & 283.9 & 8.2 & $-$269.5 & 8.2 & $-$12.7 & 4.3 & 220.0 & $-$218.2 & $-$11.5 & $-$306.4 & $-$51.8 & 283.5 & 82.9 \\
13 & 306.9 & 8.3 & $-$292.2 & 8.3 & $-$14.4 & 4.6 & 219.5 & $-$235.0 & $-$23.8 & $-$317.9 & $-$51.1 & 311.1 & 105.4 \\
14 & 329.9 & 8.4 & $-$315.0 & 8.5 & $-$16.0 & 5.0 & 218.0 & $-$249.2 & $-$34.9 & $-$326.8 & $-$50.3 & 339.6 & 130.8 \\
15 & 352.9 & 8.6 & $-$337.7 & 8.6 & $-$17.7 & 5.3 & 216.1 & $-$261.3 & $-$44.9 & $-$333.9 & $-$49.4 & 368.7 & 157.7 \\
\enddata
\tablecomments{
Column headings are:\\
1: D = Distance, kpc.  \\
2--7: U, $\sigma_{\rm U}$, V, $\sigma_{\rm V}$, W, $\sigma_{\rm W}$ = Velocities
       transformed to LSR frame from the measured proper motion and
       errors. At each distance, $\sigma_{\rm D}$ is taken as 0. E.g., At
       10~kpc, $\sigma_{\rm U}$, $\sigma_{\rm V}$, $\sigma_{\rm W}$, are 8, 8, 4
       km~s$^{-1}$, from errors on $\mu_{\rm RA}$, $\mu_{\rm DEC}$, and
       V$_{\rm Helio}$ only. \\
8,9: UC, VC = Velocities (LSR), as expected from circular rotation at
       220 km~s$^{-1}$. WC = 0.  \\
10: VR$_{\rm H}$ = Radial vel (Heliocentric), expected from circular
    rotation.  \\
11: VT$_{\rm H}$ =
    Transverse vel (Heliocentric) across line of sight, expected from
    circular rotation.  \\ 
12: VR$_{\rm G}$ = Measured radial vel
    (Galactocentric). 0 km~s$^{-1}$ expected for circular rotation.  \\ 
13: VC$_{\rm G}$ = Measured circular vel(Galactocentric). 220 km~s$^{-1}$
     expected for circular rotation.  \\ 
14: V$_{\rm P}$ = Peculiar velocity = [(U$-$UC)$^{2}$ + (V$-$VC)$^{2}$ +
    W$^{2}$]$^{0.5}$ = [VR$^{2}_{\rm G}$ + (VC$_{\rm G} - $ 220)$^{2}$ +
    W$^2$]$^{0.5}$             }

\end{deluxetable}

\clearpage
\pagestyle{empty}

\begin{figure*}[!hbt]
\epsscale{0.8}
\plotone{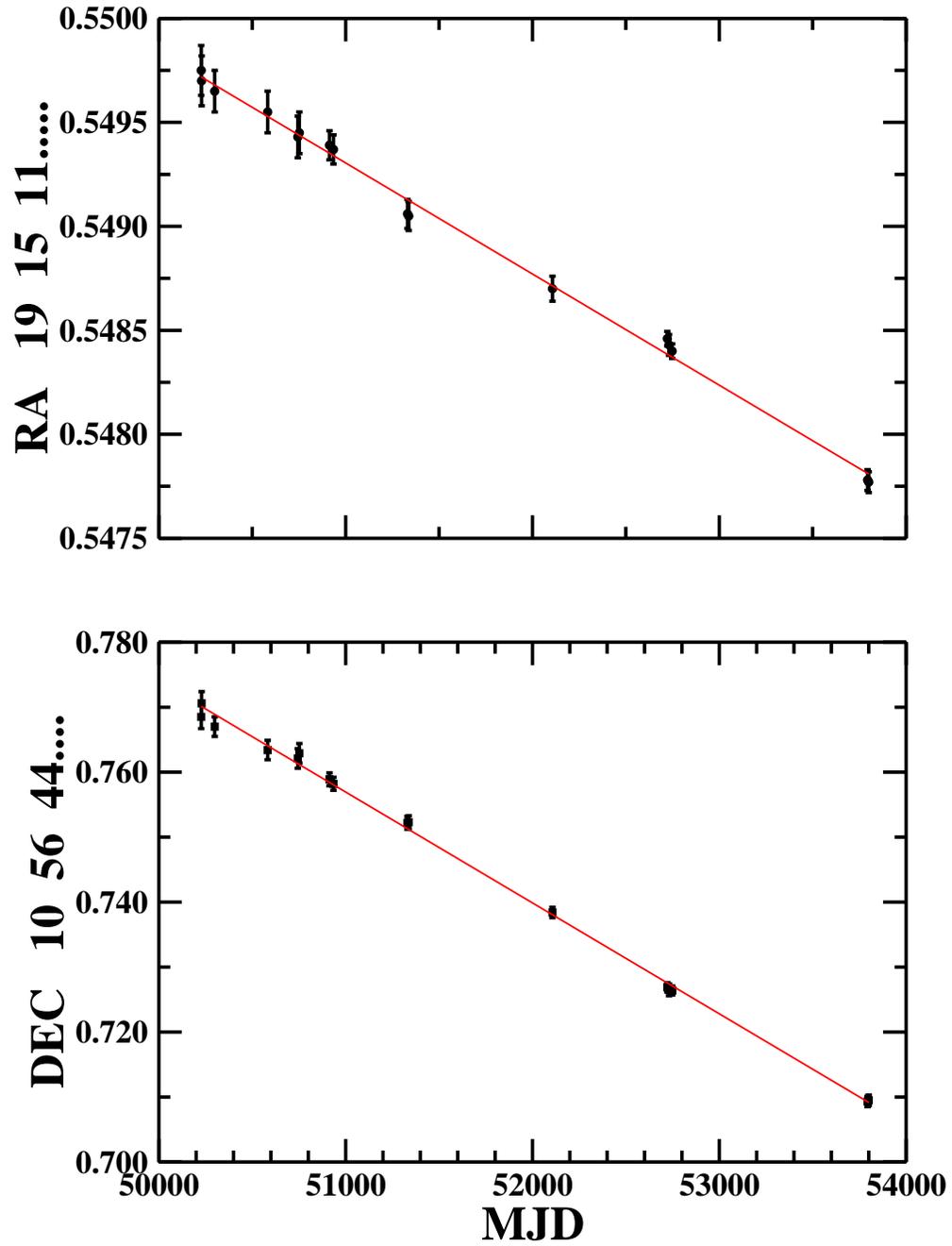}
\caption{Astrometric Position of the Black Hole over a timespan of
  almost 10 years.}
\label{fig1}
\end{figure*}

\clearpage

\begin{figure*}[!hbt]
\epsscale{0.8}
\plotone{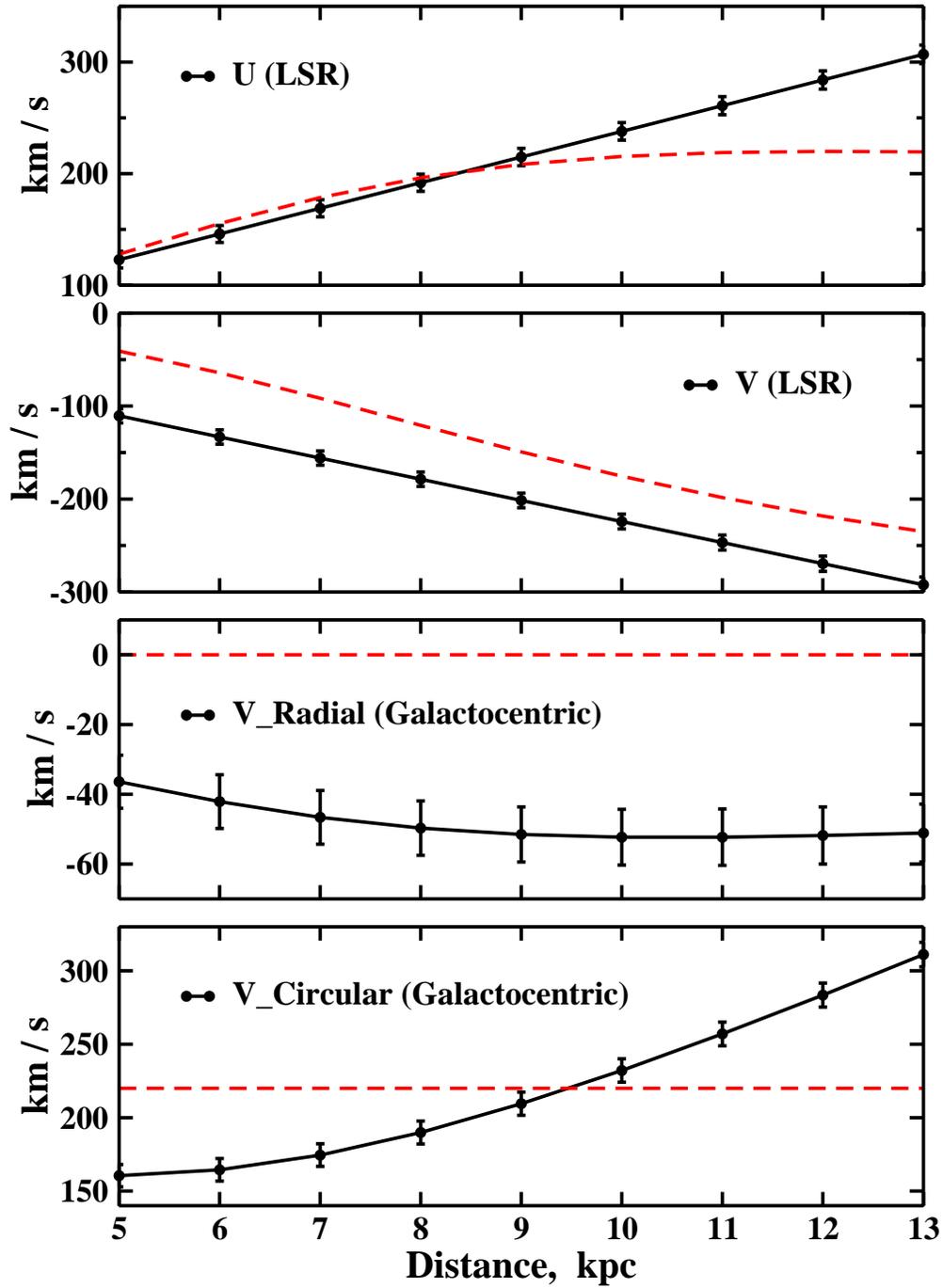}
\caption{Measured velocity components (solid lines) vs. distance,
compared with expected values for an object in circular rotation about
the Galactic Center (dashed lines).}
\label{fig2}
\end{figure*}

\clearpage

\begin{figure*}[!hbt]
\epsscale{0.9}
\plotone{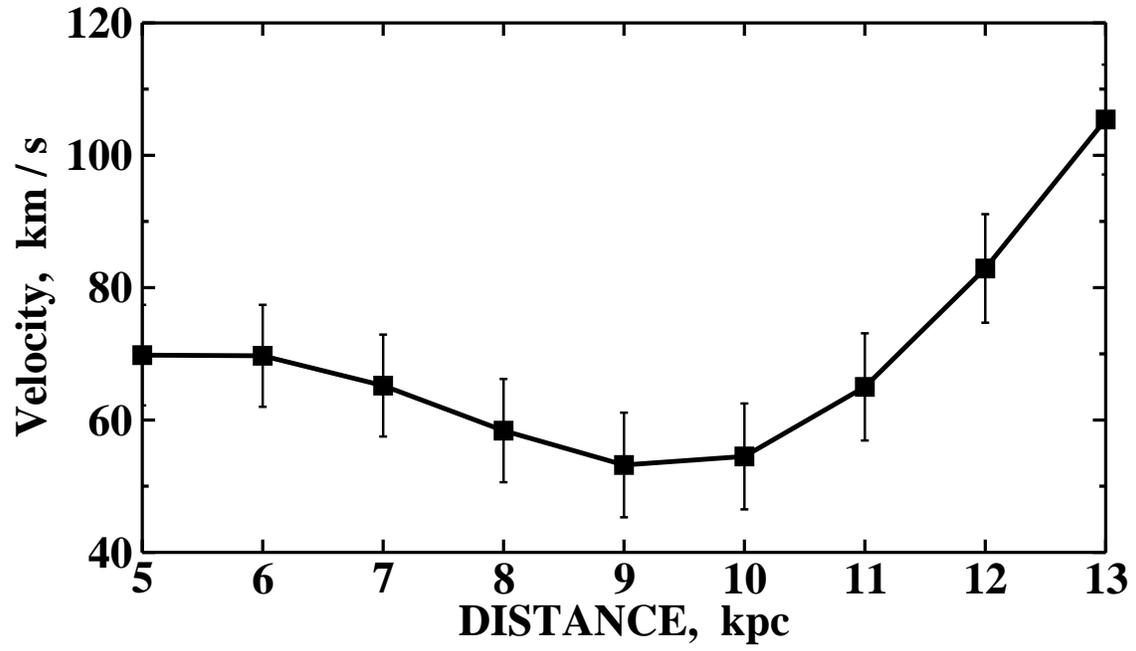}
\caption{Total peculiar velocity of GRS~1915+105 plotted at 
various assumed distances.}
\label{fig3}
\end{figure*}

\clearpage

\begin{figure*}[!hbt]
\epsscale{0.8}
\plotone{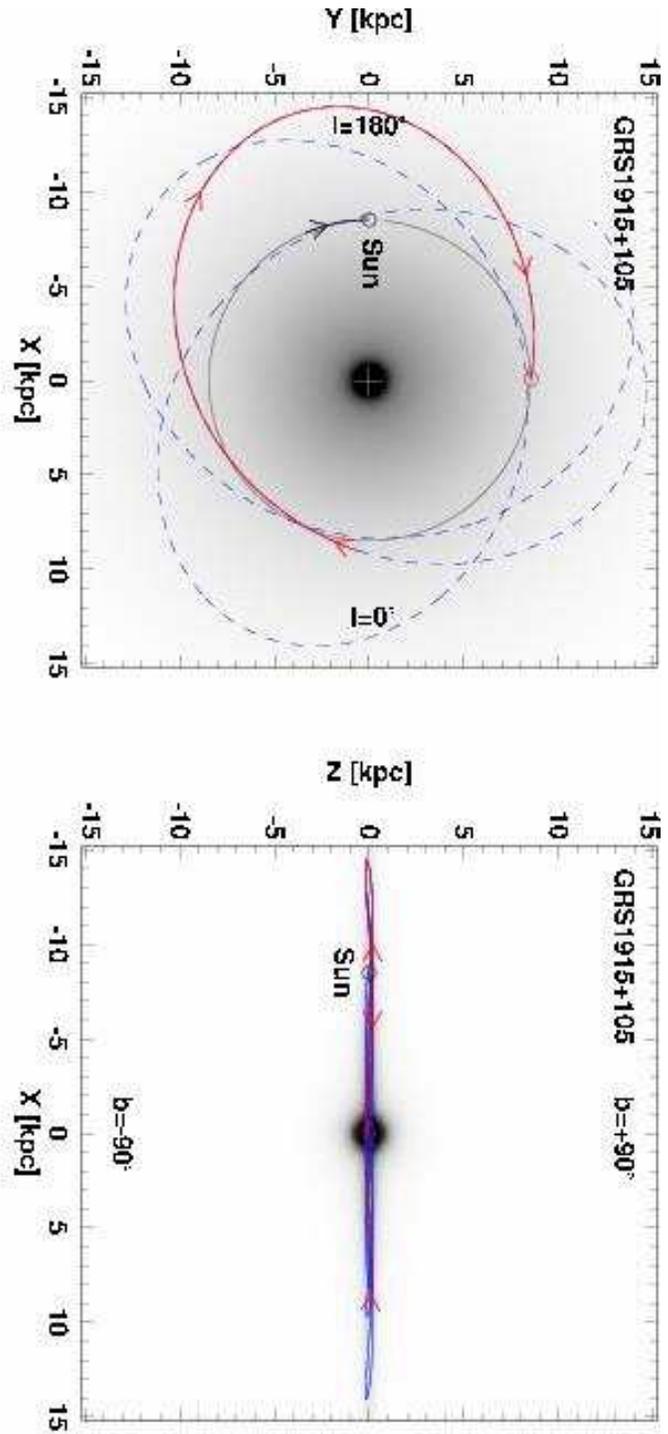}
\caption{Galactocentric Orbit of GRS~1915+105 for the past 1~Gyr,
  (dashed blue line), with the most recent 250~Myr shown as the solid
  red line.}
\label{fig4}
\end{figure*}


\begin{thebibliography}{}

\bibitem[Alcob{\'e} \& Cubarsi(2005)]{Alc05} Alcob{\'e}, S., \&
Cubarsi, R.\ 2005, \aap, 442, 929

\bibitem[Beasley \& Conway(1995)]{BC95} Beasley, A.~J., \& Conway,
J.~E.\ 1995, ASP Conf.~Ser.~ 82: Very Long Baseline Interferometry and
the VLBA, 82, 328

\bibitem[Belczynski \& Bulik(2002)]{BB02} Belczynski, K., \& Bulik,
T.\ 2002, \apjl, 574, L147

\bibitem[Belloni et al.(2000)]{Bel00} Belloni, T., 
Klein-Wolt, M., M{\'e}ndez, M., van der Klis, M., \& van Paradijs, J.\ 
2000, \aap, 355, 271 

\bibitem[Casares et al.(2005)]{Cas05} Casares, J., Rib{\'o}, M.,
Ribas, I., Paredes, J.~M., Mart{\'{\i}}, J., \& Herrero, A.\ 2005,
\mnras, 364, 899

\bibitem[Dehnen \& Binney(1998)]{DB98} Dehnen, W., \& Binney, J.~J.\
1998, \mnras, 298, 387

\bibitem[Dhawan et al.(2000a)]{Dha00a} Dhawan, V., Mirabel, I.~F., \&
Rodr{\'{\i}}guez, L.~F.\ 2000, \apj, 543, 373

\bibitem[Dhawan et al.(2000b)]{Dha00b} Dhawan, V., Goss, W.~M., \&
Rodr{\'{\i}}guez, L.~F.\ 2000, \apj, 540, 863

\bibitem[Fender(2003)]{Fen03} Fender, R.~P.\ 2003, \mnras, 340, 1353

\bibitem[Foellmi et al.(2007)]{Foe07} Foellmi, C., Dall, T.~H., \&
Depagne, E.\ 2007, \aap Letters(in press). arXiv:astro-ph/0702059

\bibitem[Fryer \& Kalogera(2001)]{Fry01} Fryer, C.~L., \& Kalogera,
V.\ 2001, \apj, 554, 548

\bibitem[Fuchs et al.(2003)]{Fuc03} Fuchs, Y., et al.\ 2003, \aap,
409, L35

\bibitem[Gonz{\'a}lez Hern{\'a}ndez et al.(2005)]{GH05} Gonz{\'a}lez
Hern{\'a}ndez, J.~I., Rebolo, R., Pe{\~n}arrubia, J., Casares, J., \&
Israelian, G.\ 2005, \aap, 435, 1185

\bibitem[Gelino et al.(2006)]{Gel06} Gelino, D.~M., Balman, 
{\c S}., K{\i}z{\i}lo{\u g}lu, {\"U}., Y{\i}lmaz, A., Kalemci, E., \& 
Tomsick, J.~A.\ 2006, \apj, 642, 438 

\bibitem[Gonz{\'a}lez Hern{\'a}ndez et al.(2006)]{Gon06} Gonz{\'a}lez
Hern{\'a}ndez, J.~I., Rebolo, R., Israelian, G., Harlaftis, E.~T.,
Filippenko, A.~V., \& Chornock, R.\ 2006, \apjl, 644, L49

\bibitem[Gonz{\'a}lez Hern{\'a}ndez et al.(2007)]{GH07} Gonz{\'a}lez
Hern{\'a}ndez, J.~I., et al. 2007, \aap, submitted.

\bibitem[Greene et al.(2001)]{Gre01} Greene, J., Bailyn, 
C.~D., \& Orosz, J.~A.\ 2001, \apj, 554, 1290 

\bibitem[Greiner et al.(1996)]{GMR96} Greiner, J., Morgan, E.~H., \&
Remillard, R.~A.\ 1996, \apjl, 473, L107

\bibitem[Greiner et al.(2001)]{GCM01} Greiner, J., Cuby, J.~G., \&
McCaughrean, M.~J.\ 2001, \nat, 414, 522

\bibitem[Hobbs et al.(2005)]{Hob05} Hobbs, G., Lorimer, D.~R., Lyne,
A.~G., \& Kramer, M.\ 2005, \mnras, 360, 974

\bibitem[Israelian et al.(1999)]{Isr99} Israelian, G., Rebolo, R.,
Basri, G., Casares, J., \& Mart{\'{\i}}n, E.~L.\ 1999, \nat, 401, 142

\bibitem[Johnson \& Soderblom(1987)]{JS87} Johnson, D.~R.~H., \&
Soderblom, D.~R.\ 1987, \aj, 93, 864

\bibitem[Johnston et al.(1995)]{JSH95} Johnston, K.~V.,
Spergel, D.~N., \& Hernquist, L.\ 1995, \apj, 451, 598

\bibitem[Kaiser et al.(2004)]{Kai04} Kaiser, C.~R., Gunn, K.~F.,
Brocksopp, C., \& Sokoloski, J.~L.\ 2004, \apj, 612, 332

\bibitem[McSwain et al.(2004)]{McS04} McSwain, M.~V., Gies, D.~R.,
Huang, W., Wiita, P.~J., Wingert, D.~W., \& Kaper, L.\ 2004, \apj,
600, 927

\bibitem[Mignard(2000)]{Mig00} Mignard, F.\ 2000, \aap, 354, 522
 
\bibitem[Miller-Jones et al.(2007)]{MJ07} Miller-Jones, J.~C.~A.,
Rupen, M.~P., Fender, R.~P., Rushton, A., Pooley, G.~G., \& Spencer,
R.~E.\ 2007, \mnras, 375, 1087

\bibitem[Mirabel \& Rodriguez(1994)]{MR94} Mirabel, I.~F., \&
Rodriguez, L.~F.\ 1994, \nat, 371, 46

\bibitem[Mirabel et al.(1998)]{Metal98} Mirabel, I.~F., Dhawan, V.,
Chaty, S., Rodriguez, L.~F., Marti, J., Robinson, C.~R., Swank, J., \&
Geballe, T.\ 1998, \aap, 330, L9

\bibitem[Mirabel et al.(2001)]{Metal01} Mirabel, I.~F., Dhawan, V.,
Mignani, R.~P., Rodrigues, I., \& Guglielmetti, F.\ 2001, \nat, 413,
139

\bibitem[Mirabel et al.(2002)]{Metal02} Mirabel, I.~F., Mignani, R.,
Rodrigues, I., Combi, J.~A., Rodr{\'{\i}}guez, L.~F., \& Guglielmetti,
F.\ 2002, \aap, 395, 595

\bibitem[Mirabel \& Rodrigues(2003a)]{MR03a} Mirabel, I.~F., \&
Rodrigues, I.\ 2003, Science, 300, 1119

\bibitem[Nelemans et al.(1999)]{Nel99} Nelemans, G., Tauris, T.~M., \&
van den Heuvel, E.~P.~J.\ 1999, \aap, 352, L87

\bibitem[Pfahl et al.(2002)]{Pfa02} Pfahl, E., Rappaport, S., 
Podsiadlowski, P., \& Spruit, H.\ 2002, \apj, 574, 364 

\bibitem[Rib{\'o} et al.(2002)]{Rib02} Rib{\'o}, M., Paredes, J.~M.,
Romero, G.~E., Benaglia, P., Mart{\'{\i}}, J., Fors, O., \&
Garc{\'{\i}}a-S{\'a}nchez, J.\ 2002, \aap, 384, 954

\bibitem[Rib{\'o} et al.(2004)]{RDM04} Rib{\'o}, M., Dhawan, V., \&
Mirabel, I.~F.\ 2004, European VLBI Network on New Developments in
VLBI Science and Technology, 111

\bibitem[Sellwood \& Binney(2002)]{SB02} Sellwood, J.~A., \& Binney,
J.~J.\ 2002, \mnras, 336, 785

\bibitem[Weilen(1977)]{Wei77} Weilen, R., 1977, \aap, 60, 263

\bibitem[Willems et al.(2005)]{Wil05} Willems, B., Henninger, M.,
Levin, T., Ivanova, N., Kalogera, V., McGhee, K., Timmes, F.~X., \&
Fryer, C.~L.\ 2005, \apj, 625, 324

\end{thebibliography}
\end{document}